\documentclass{article}

\usepackage{amsmath}
\usepackage{amsthm}
\usepackage{color}
\usepackage{graphicx}
\usepackage{picture}
\usepackage{url}
\usepackage{subfig}
\usepackage{ams}

\begin{document}

\title{RS-232 LED Board Project}
\author{By Vladimir Tskhvaradze \\ \textcolor{blue}{\small{vtms1@yahoo.com}}}
\maketitle

\begin{abstract}
This article demonstrates how to develop a Microchip PIC16F84 based
device that supports RS-232 interface with PC. Circuit (LED Board)
design and software development will be discussed. PicBasic Pro
Compiler from microEngineering Labs, Inc. is used for PIC
programming. Development of LED Board Control Console using C/C++ is
also briefly discussed. The project requires basic work experience
with Microchip PICs, serial communication and programming.
\\
\\
\\
\begin{center}
\includegraphics[width=2in]{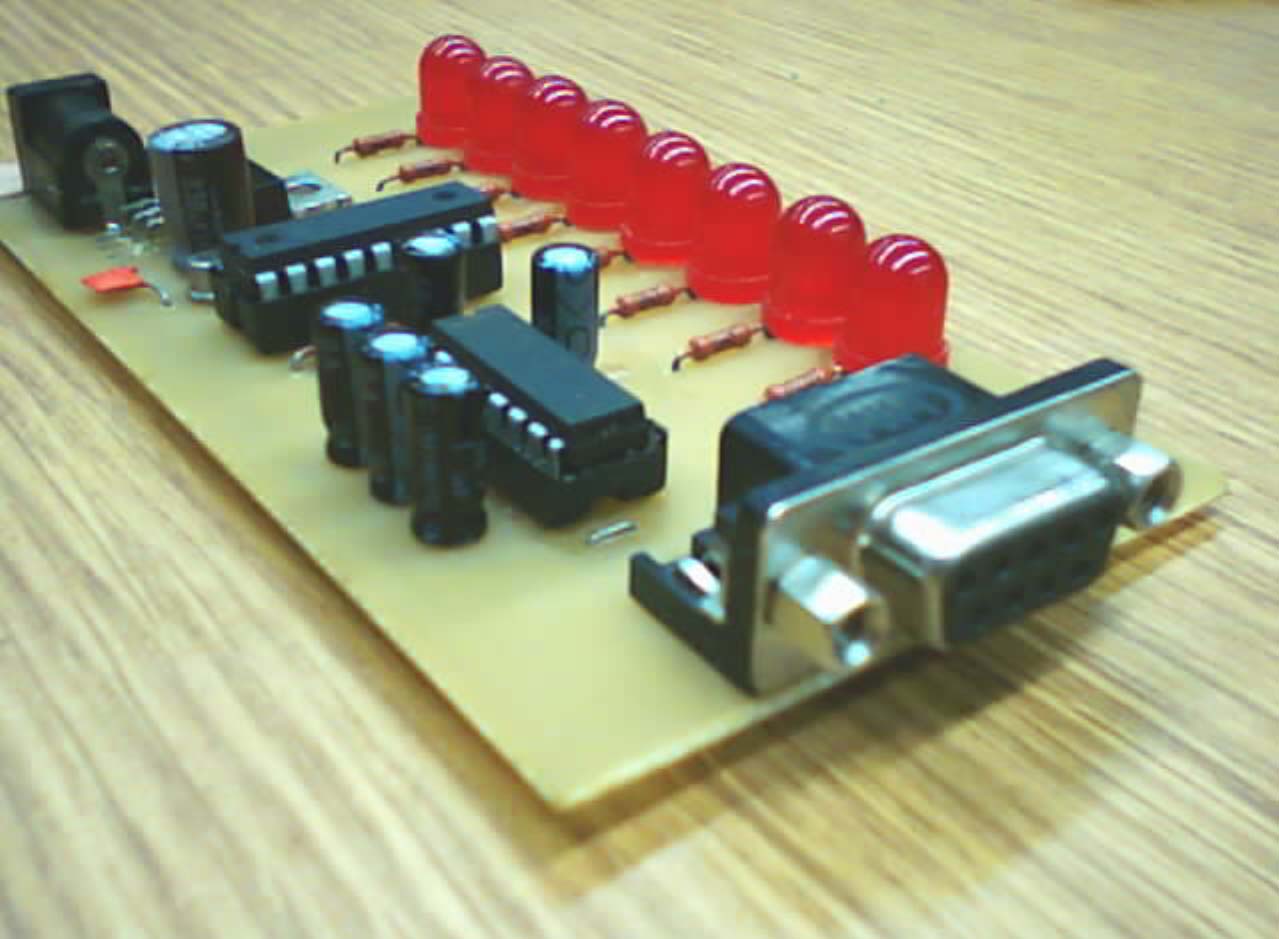}
\end{center}
\end{abstract}

\thispagestyle{empty}

\newpage

\section{Introduction}

LED Board is a printed circuit board (PCB) that connects to PC via
standard serial cable. LED Board has 8 LEDs, each LED can be
individually turned On/Off from Windows application (LED Board
Control Console). Fig.1 shows simplified schematic of LED Board.
Microchip PIC16F84 is responsible for receiving data from PC and
controlling 8 LEDs.

\makeatletter
\def\fps@figure{h}
\makeatother

\begin{figure}
\centering
\includegraphics[totalheight=2.45in]{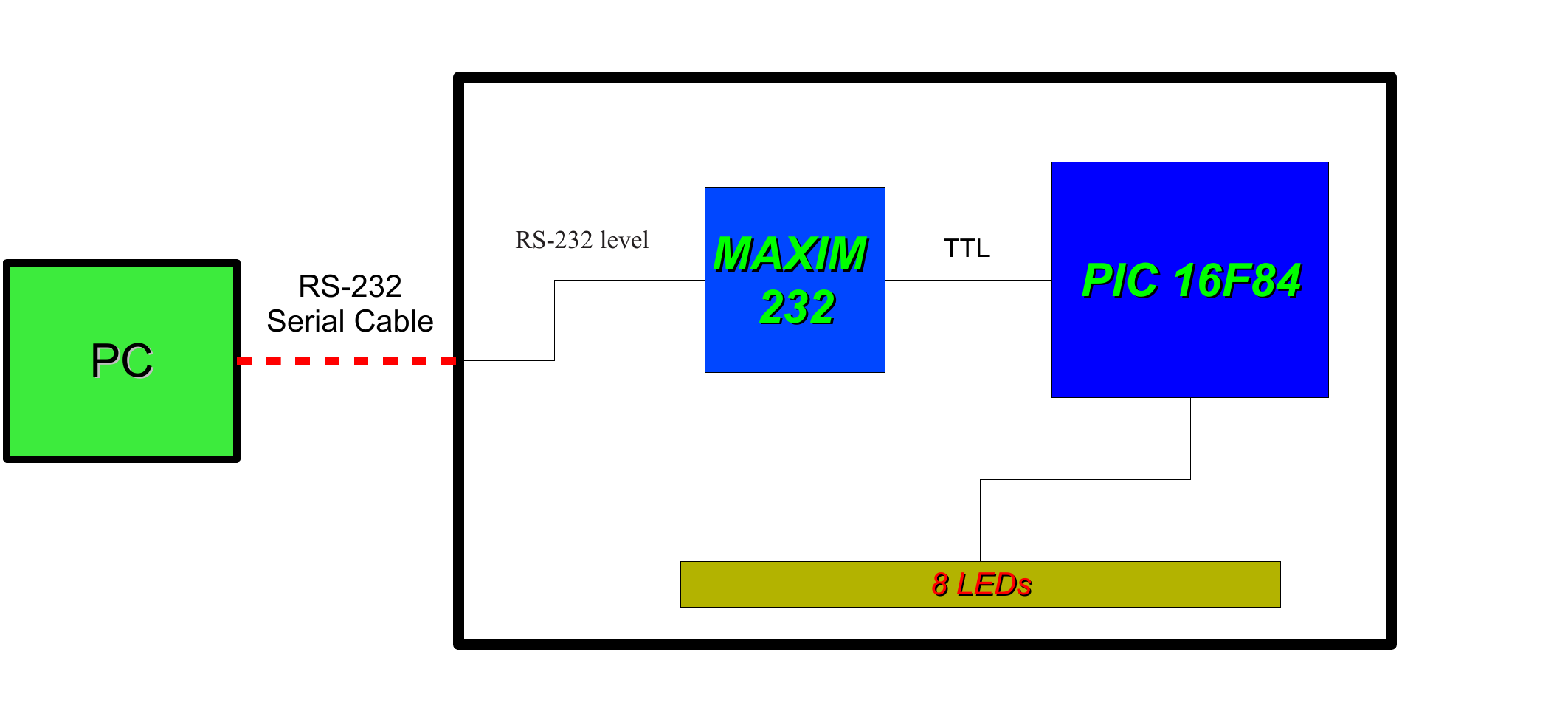}
\caption{LED Board schematic} \label{fig:graph}
\end{figure}
Since RS-232 and PIC16F84 use different voltage levels, MAX232 is
required for logic level conversion: RS-232 $\longleftrightarrow$
TTL. The Null Modem cable configuration without handshaking is used
between PC and LED Board.

%------------------------------------------------------------------------------------------------------------------
%   SCHEMATIC
%------------------------------------------------------------------------------------------------------------------
\newpage
\section{LED Board schematic}

\begin{center}
\includegraphics[width=7.15in, angle=90]{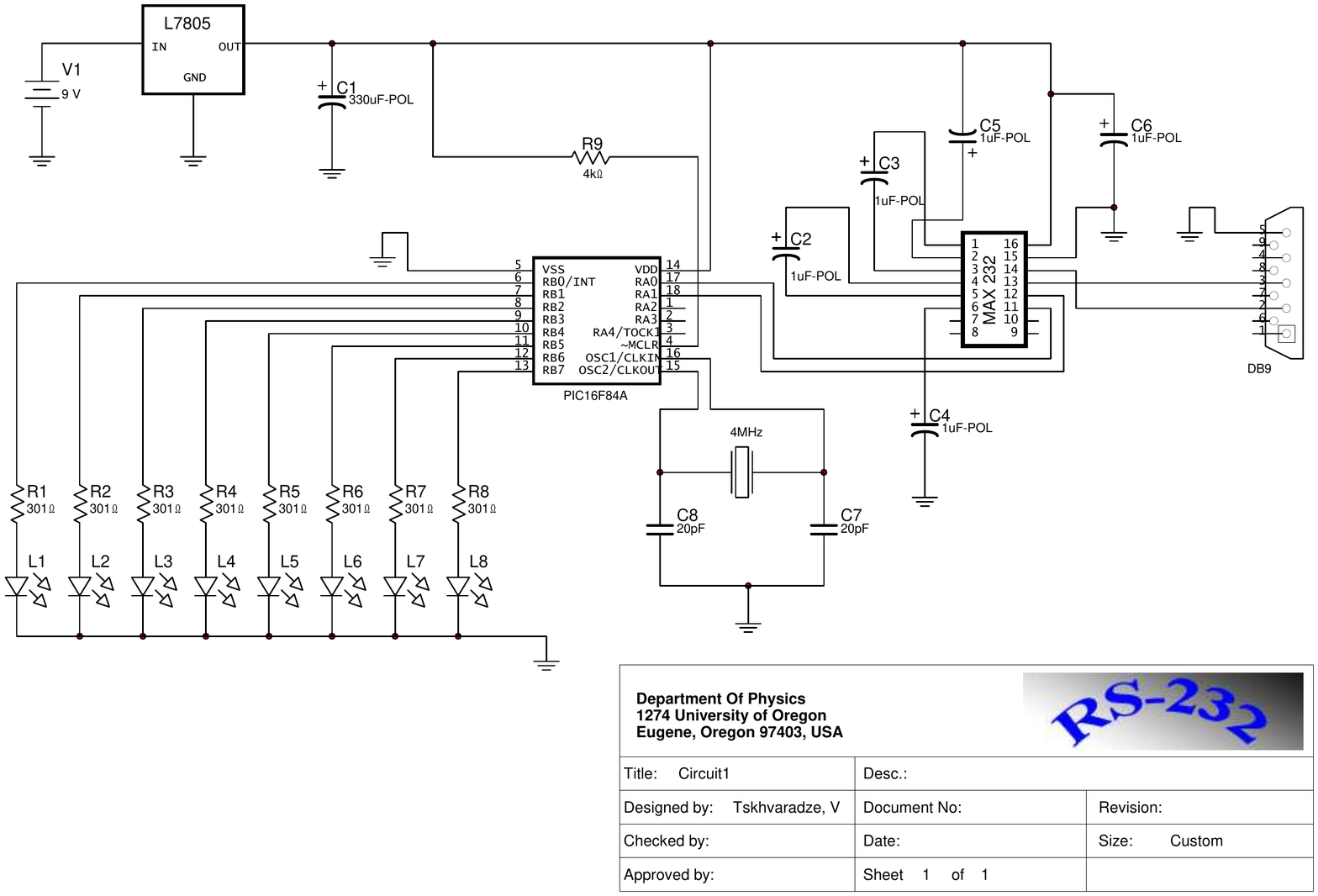}
\end{center}

%------------------------------------------------------------------------------------------------------------------
%   SCHEMATIC DESCRIPTION
%------------------------------------------------------------------------------------------------------------------
\newpage

\section{LED Board schematic description}
As we can see from the schematic MAX232 is connected to Microchip
PIC 16F84 and DB9 serial connector. To provide +5V for PIC16F84 and
MAX232, we use 9V battery together with L7805 voltage regulator.
LEDs are directly connected to the PIC16F84's PORTB I/O pins which
have up to 25 mA sink/source current, this feature makes them ideal
for direct LED drive. 4MHz crystal resonator and two 20pF capacitors
(C7, C8) are used to provide external clocking (XT mode). The Null
Modem configuration uses TxD and RxD lines for data interchange
between PC and external device. LED Board uses TxD to receive data,
RxD is wired-up, but not used. TxD is connected to MAX232's R1IN
(pin 13), while R1OUT (pin 12) is directly connected to PIC16F84's
RA1 (pin 18), thus RA1 is used as an input pin which receives data
from PC\footnote{There is nothing special with RA1, different pin
can be used as well}. Please see datasheets of Microchip PIC16F84,
MAX232, L7805 for detailed specifications.

%------------------------------------------------------------------------------------------------------------------
%   PICBASIC CODE
%------------------------------------------------------------------------------------------------------------------
\section{PicBasic Pro code for Microchip PIC 16F84}
\vspace*{0.1in} \fbox{\begin{tabular}{p{5in}}
\emph{; LED control program by Vladimir Tskhvaradze}\\\\

Include "modedefs.bas"    ~~~~~~~~\emph{;Include serial modes}\\\\

pinin var PORTA.1     ~~~~~~~~~~~~\emph{;Define pinin as PORTA.1}\\
B0    var byte\\\\

POKE TRISB,0  ~~~~~~~\emph{;Set PORTB to output}\\
POKE TRISA,31  ~~~~~\emph{;Set PORTA to input}\\
POKE PORTB,0   ~~~~~\emph{;Turn off all LEDs}\\\\

Loop:\\

\hspace*{0.51in} SERIN pinin,T2400,B0\\
\hspace*{0.51in} POKE PORTB,B0\\

GOTO Loop\\

\end{tabular}}
\\ \\ \\
Above program receives bytes and depending on bit pattern of each
byte switches corresponding LEDs On/Off. For example, if received
byte is 00010100, LED \#3 and LED \#5 will be switched On, while
remaining LEDs will be Off (1 = LED ON, 0= LED OFF). Obviously, to
light up all 8 LEDs, PIC16F84 should receive 11111111 . After SERIN
receives a byte, it stores received byte in B0. Once byte has been
received, it is written into PORTB with POKE command, thus switching
LEDs On/Off according to above mentioned bit pattern. Since we have
exactly 8 LEDs connected to PORTB (8 pins), it is easy to control
them all with just one byte.
%------------------------------------------------------------------------------------------------------------------
%   Microsoft Visual C++ code
%------------------------------------------------------------------------------------------------------------------
\section{LED Board Control Console application}
LED Board Control Console (LBCC) is a small Windows application
designed to control LED Board from PC. To switch LED(s) On/Off, LBCC
sends a byte with specified bit pattern into COM port. Fig.2 shows
sample application developed with Microsoft's Visual C++.

\vspace*{0.1in}

\begin{figure}
\centering
\includegraphics[totalheight=2.25in]{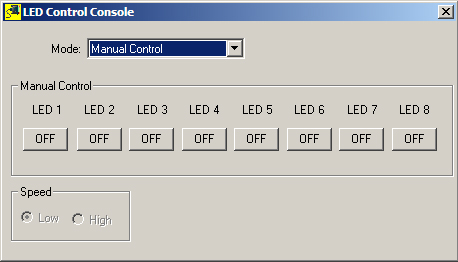}
\caption{LED Board Control Console}{\hspace*{0.86in}(Dialog based
MFC application)} \label{fig:graph}
\end{figure}

\subsection{COM port setup}
Before sending data, we should setup and configure corresponding COM
port. WinAPI function \emph{CreateFile (\ldots)} creates serial
port, it returns handle which is later used to access serial port.
The following code shows how to successfully setup COM
port\footnote{To setup serial port other than the COM1, just pass
different parameter to \emph{CreateFile} e.g. "COM2"}.
\begin{flushleft}
\verb"HANDLE  ComPort;     // COM port handle"\\
\verb"DCB dcb;             // Control structure for a serial device"
\verb"BOOL fSuccess;"\\
\end{flushleft}
\vdots
\begin{verbatim}
ComPort = CreateFile("COM1",GENERIC_READ | GENERIC_WRITE,
                      0,            // comm devices must be
                                    // opened w/exclusive-access
                     NULL,          // no security attributes
                     OPEN_EXISTING, // comm devices
                                    // must use OPEN_EXISTING
                     0,             // not overlapped I/O
                     NULL           // hTemplate must be NULL
                                    //for comm devices
                     );

if (ComPort == INVALID_HANDLE_VALUE)
  {
      // Handle the error.
      MessageBox ("CreateFile failed","ERROR",MB_OK);
      return (1);
  }

  fSuccess = GetCommState(ComPort, &dcb);

  if (!fSuccess)
  {
      // Handle the error.
      MessageBox ("GetCommState failed","ERROR",MB_OK);
      return (2);
  }

  // Filling in DCB structure
  //
  // IMPORTANT NOTE:
  // Serial device configuration (e.g. speed, parity) should be matched with
  // the corresponding SERIN mode (See section 4 for PICBASIC code)

  dcb.BaudRate = CBR_2400;      // set the baud rate
  dcb.ByteSize = 8;             // data size, xmit, and rcv
  dcb.Parity = NOPARITY;        // no parity bit
  dcb.StopBits = ONESTOPBIT;    // one stop bit

  fSuccess = SetCommState(ComPort, &dcb);

  if (!fSuccess)
  {
      // Handle the error.
      MessageBox ("SetCommState failed","ERROR",MB_OK);
      return (3);
    }

  MessageBox ("COM port initialized",NULL,MB_OK);
\end{verbatim}

COM port is now initialized and ready for use.

\subsection{Sending data to LED Board}

To transmit data to LED Board, we use WinAPI function
\emph{WriteFile(\ldots)} which writes a byte into COM port. As an
example, let's consider the following, suppose we want to switch on
LED \#3 and LED \#5 while keeping remaining LEDs Off, this can be
done by sending the byte 00010100 into COM port.

\begin{verbatim}
__int8 ComByte;

DWORD nNumberOfBytesToWrite = 1;
DWORD NumberOfBytesWritten;

ComByte = 20;  // Binary 00010100 corresponds to 20 (decimal)

WriteFile(ComPort,&ComByte,nNumberOfBytesToWrite,&NumberOfBytesWritten,NULL);
\end{verbatim}
The first parameter in \emph{WriteFile(\ldots)} is COM port handle
(section 5.1).
\subsection{Switching On/Off specified LED.\\ Bitwise operations.}
Suppose we need to switch On specified LED without affecting other
LEDs (if some of them were On (Off), they should remain On (Off)).
Let's also assume that the LED we want to switch On was previously
Off. To solve this task we will use successive bitwise operations.
Assume that we have the following bit pattern 00010010 (LED \#2 is
On, LED \#5 is On; remaining LEDs are Off) and we want to switch On
LED \#7. Let's use OR mask with the original bit pattern (00010010)
\begin {verbatim}
__int8 ORMask = 1;
__int8 ComByte = 18;  // 00010010
\end{verbatim}
\verb"ORMask = ORMask << 6; // 00000001" $\longrightarrow$
\verb"01000000"\\
\verb"ComByte = ComByte | ORMask; // 00010010 | 01000000 = 01010010"

\vspace*{0.4cm}

ComByte can now be written to COM port with \emph{WriteFile(\ldots)}
As we can see from bit pattern LED \#7 is switched On, while other
LEDs are not affected. Please note that to switch On/Off $N^{th}$
LED, left shift should be applied $N-1$ times. In our case $N=7$,
therefore we have ORMask = ORMask $<<$ 6. Analogously, to switch Off
LED \#7, XOR mask should be used

\begin{verbatim}
__int8 XORMask = 1;
__int8 ComByte = 82;  // 01010010
\end{verbatim}
\verb"XORMask = XORMask << 6; // 00000001" $\longrightarrow$
\verb"01000000"\\
\verb"ComByte = ComByte ^ XORMask; //01010010 ^ 01000000 = 00010010"

\section{LED Board PCB}
PCB size (Approx.):\\
Length: \emph{112.5 mm (4.43 in)}\\
Width: \emph{57 mm (2.24 in)}\\

%------------------------------------------------------------------------------------------------------------------
%   PCB figures (left & right)
%------------------------------------------------------------------------------------------------------------------
\begin{figure}

%%----start of first subfigure----
\subfloat[PCB screenshot]{
\label{fig:subfig:a} %% label for first subfigure
\includegraphics[width=2.0in]{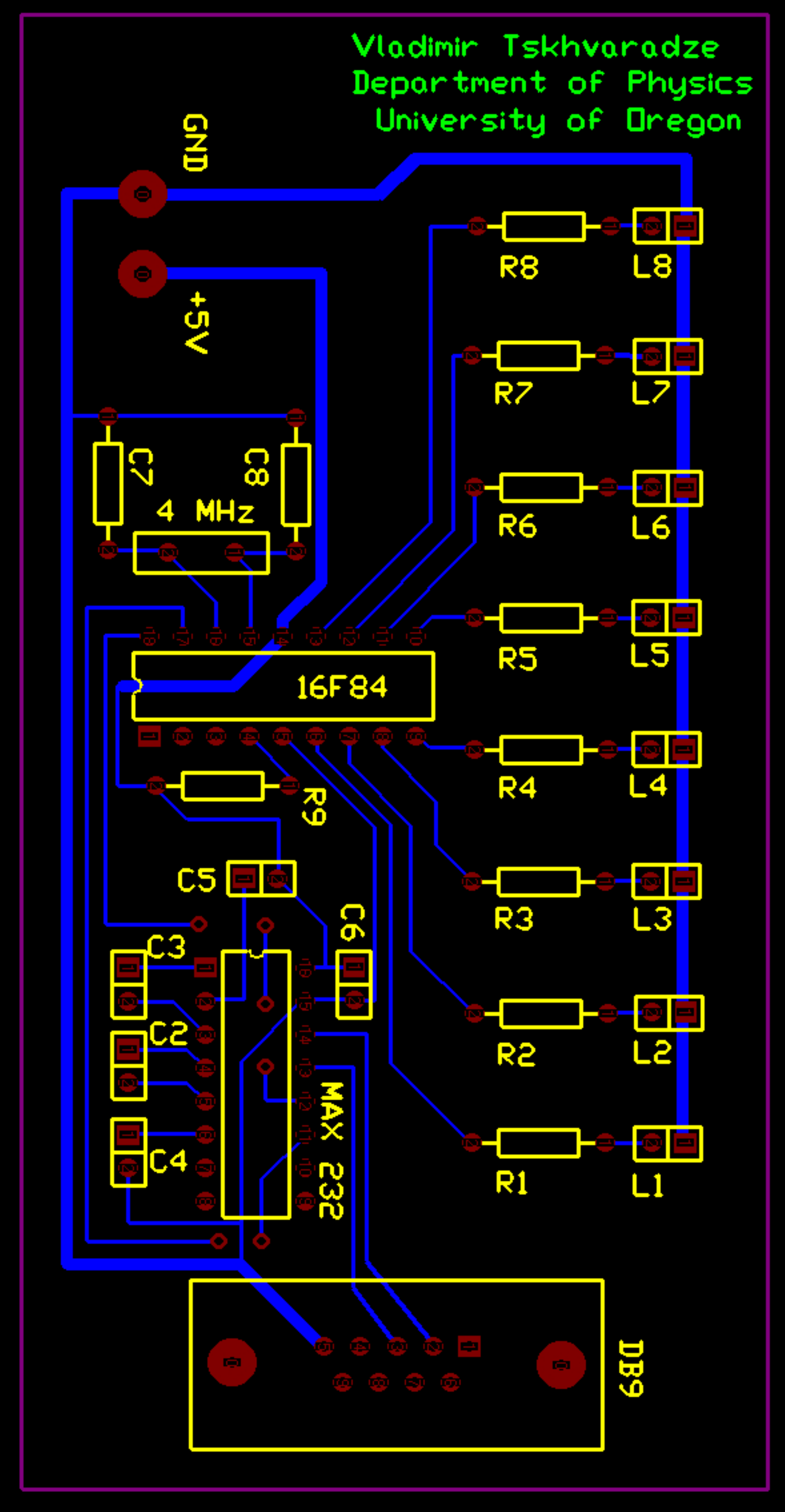}}
\hspace{1in}
%%----start of second subfigure----
\subfloat[PCB without top overlay]{
\label{fig:subfig:b} %% label for second subfigure
\includegraphics[width=2.5in]{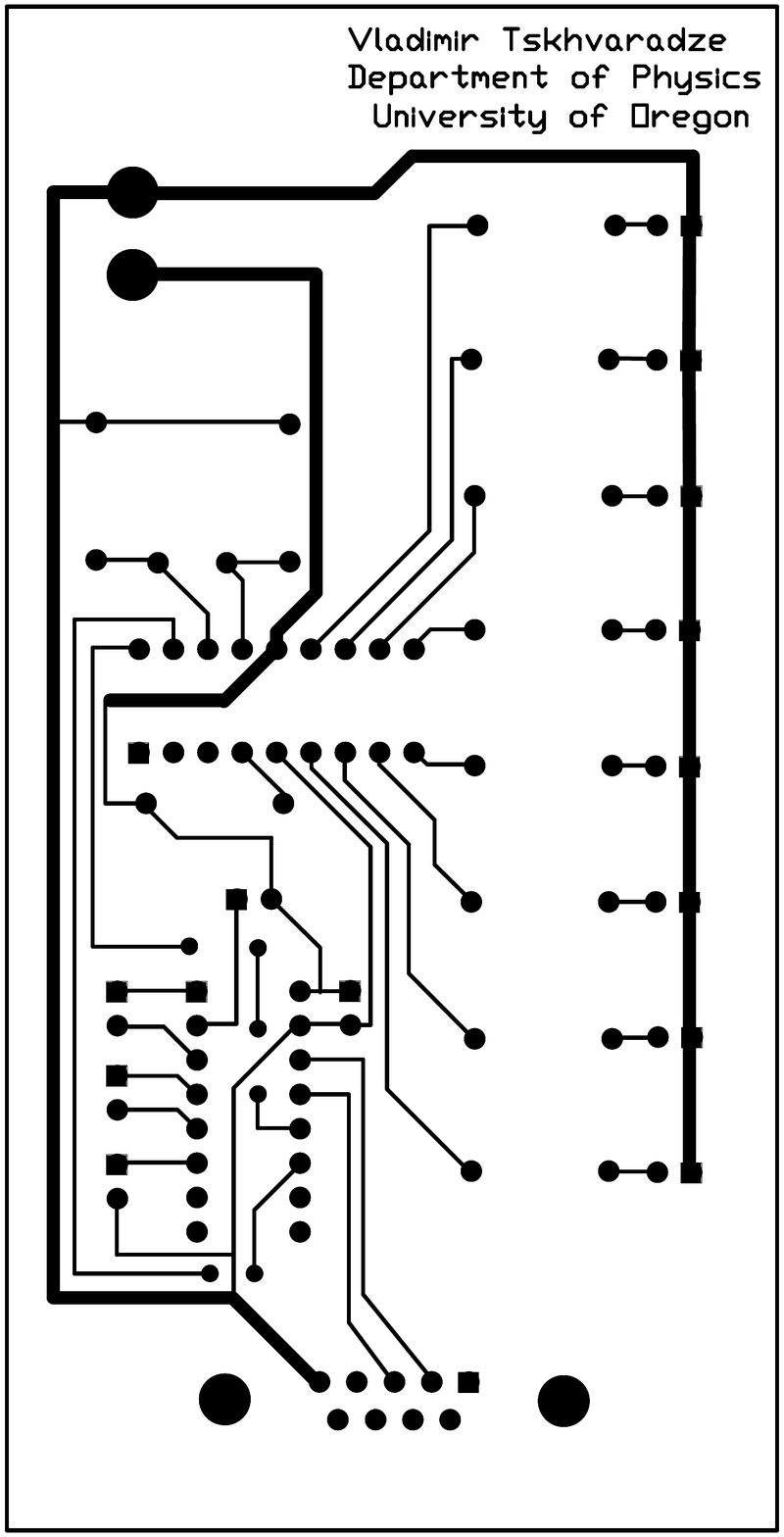}}
\caption{LED Board PCB (L7805 not included)}
\label{fig:subfig} %% label for entire figure
\end{figure}

%------------------------------------------------------------------------------------------------------------------
%   Parts List
%------------------------------------------------------------------------------------------------------------------
\section{Parts List}
\begin{flushleft}
\begin{tabular}{|c|c|}
\hline
\textcolor{blue}{\textbf{Part}} & \textcolor{blue}{\textbf{Quantity}}\\
\hline
Microchip PIC16F84 & 1 \\
\hline
MAX232 (+5V Powered RS-232 Driver/Receiver) & 1\\
\hline
ST L7805 (Positive Voltage Regulator (TO-220)) & 1 \\
\hline
Crystal Resonator (4 MHz) & 1 \\
\hline
DB-9 Connector & 1 \\
\hline
LED (Red) & 8 \\
\hline
301$\Omega$ resistor & 8 \\
\hline
4k$\Omega$ resistor & 1 \\
\hline
1$\mu$F, 50V Capacitor & 5 \\
\hline
330$\mu$F, 16V Capacitor & 1 \\
\hline
20pF, 16V Capacitor & 2 \\
\hline
DC power jack & 1 \\
\hline
\end{tabular}
\end{flushleft}

\section{Conclusion}
RS-232 LED Board project discussed in this article shows the basics
of serial communications. This project was the partial requirement
for Advanced Digital Electronics course taken at the University of
Oregon, Eugene, USA.

\vspace*{1cm}
\begin{center}
\includegraphics[width=4in]{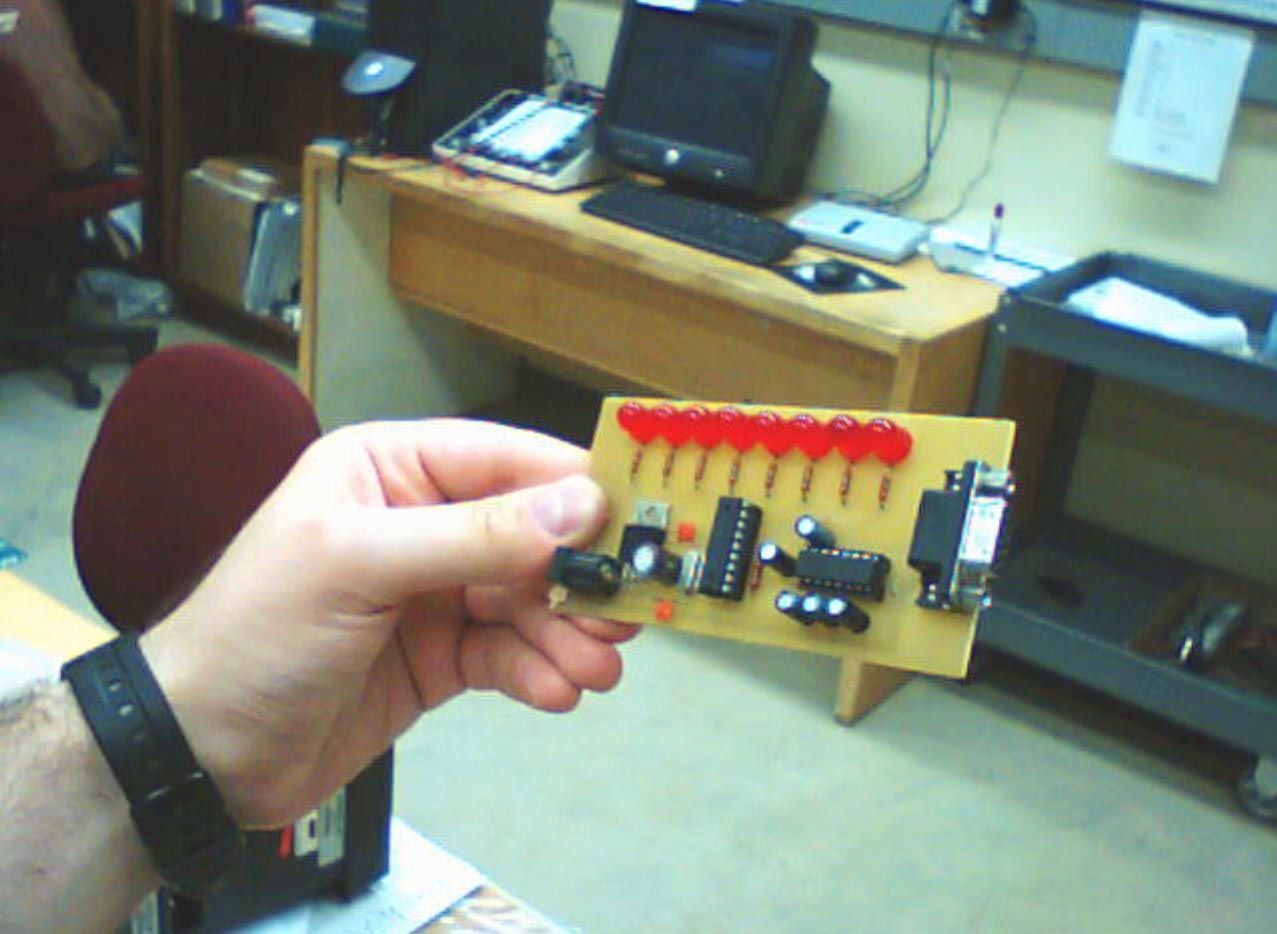}
\end{center}

\end{document}